\documentclass[twocolumn]{autart}
\usepackage{cite}
\usepackage{amsmath,amssymb,amsfonts}
\usepackage{graphicx}
\usepackage{savesym}
\savesymbol{AND}
\usepackage{algorithm, algorithmic}
\usepackage{textcomp}
\usepackage{mathabx}
\usepackage{xcolor}
\usepackage{mathtools}
\usepackage{bm}
 \usepackage{wrapfig}

\newcommand{\hpb}{h_{\mathrm{PB}}}

\newcommand{\Spb}{\mathcal{S}_{\mathrm{PB}}}
\newcommand{\Dpb}{\mathcal{D}_{\mathrm{PB}}}

\newtheorem{theorem}{Theorem}

\newtheorem{definition}{Definition}
\newtheorem{assumption}{Assumption}
\newtheorem{remark}{Remark}

\begin{document}
\begin{frontmatter}

\title{A multiobjective approach to robust predictive control barrier functions for discrete-time systems} 


\author[1]{Alexandre Didier}\ead{adidier@ethz.ch},    
\author[1]{Melanie N. Zeilinger}\ead{mzeilinger@ethz.ch}               

\address[1]{Institute for Dynamic Systems and Control, ETH Zurich, 8092 Zurich, Switzerland}  

\begin{abstract}                          
We present an optimisation-based approach to ensure robust asymptotic stability stability of a desired set in the state space of nonlinear dynamical systems, while optimising a general control objective. The approach relies on the decrease of a robust predictive control barrier function (PCBF), which is defined as the optimal value function of a slack minimisation problem with respect to the target set. We show continuity of the proposed robust PCBF, allowing the introduction of a decrease constraint in the control objective minimisation. The PCBF decrease is given with respect to a warmstart value based on a feasible solution at the prior time step. Thereby, the control objective can be optimised while ensuring robust asymptotic stability of the target set. We demonstrate the effectiveness of the proposed formulation on a linear space rendezvous and nonlinear lane changing problem.
\end{abstract}

\end{frontmatter}

\section{Introduction}
\label{sec:introduction}
Predictive control methods have become a popular tool to tackle complex control tasks, such as, e.g., quadrotor racing in \cite{krinnermpcc++} 
or the control of quadrupedal robots in \cite{neunert2018whole}. Model predictive control (MPC), see, e.g. \cite{rawlings2017} for an overview, typically makes use of a receding horizon formulation, where given a new state measurement, an optimisation problem is solved to compute an input by planning over a fixed time horizon.
The optimisation-based formulation of predictive control methods allows for incorporating constraints on system states and inputs in a natural manner over the planning horizon, while minimising a cost of a given objective. Through principled analysis, theoretical guarantees on the proposed predictive control methods can be established, such as, e.g., stability of a desired setpoint and closed-loop constraint satisfaction.

Another application of predictive control methods is the predictive safety filter, see, e.g. \cite{wabersich2023} for an overview. The aim is to augment any controller such that constraint satisfaction can be guaranteed through MPC-like arguments. The approach is agnostic to the controller which is used, such that the framework is applicable to learning-based control methods or human operators by minimising the distance of the proposed input to the input which is applied to the system. 

However, for many applications pure safety-ensuring interventions can overly compromise performance and cost functions designed to stabilise a setpoint can neglect other application-relevant requirements. A car performing a lane changing manoeuvre should reach the desired lane while, e.g., optimising passenger comfort and a satellite should reach its planned orbit while minimising fuel consumption. Additionally, predictive control methods rely on a prediction model which is used to establish theoretical guarantees. As models typically are never entirely accurate, the need arises to consider model uncertainties, such as exogenous disturbances in the optimisation problem formulation and theoretical analysis.

\textit{Contributions.} 
 In this work, we aim to address the problem of optimising a primary control objective while guaranteeing robust asymptotic stability of a target set with respect to bounded exogenous disturbances in the state space in a multiobjective approach. First, we propose a robust predictive control barrier function (PCBF) approach augmenting the nominal formulation originally proposed in \cite{wabersich2022}. The PCBF method in \cite{wabersich2022} allows for the stabilisation of a desired target set in the state space by using a slacked constraint formulation. It relies on the solution of two separate optimisation problems, i.e. minimising constraint violations in terms of the slack variables in a first step and then optimising the primary control objective given these optimal slack variables. By employing a robust positively invariant (RPI) set in the design of the optimisation problem, robust stability guarantees of the target set are established and continuity of the optimal value function of the slack minimisation problem is shown. This fact is then leveraged to propose a single optimisation problem which directly minimises the primary objective and retains stability guarantees through a decrease constraint of the PCBF value. The proposed formulation is shown to lead to less restrictive stability requirements in closed-loop compared to \cite{wabersich2022} through the introduction of a tunable hyperparameter for the PCBF decrease rate, allowing for improved primary objective optimisation. The multiobjective PCBF approach extends initial results in \cite{didier2024predictive}, where a Lyapunov decrease constraint is used to guarantee robust stability with respect to a setpoint, which may be unnecessarily restrictive in, e.g., a lane changing example for an automotive application.
 We demonstrate the effectiveness of the proposed method in simulation on a space rendezvous and lane changing example.

\textit{Related work.}
The PCBF algorithm was originally proposed in \cite{wabersich2022} and ensures asymptotic stability of the implicitly defined safe set of a predictive optimisation problem. Recently, it was shown in \cite{huang2025predictive} that a slightly modified PCBF optimization problem guarantees convergence of the system to the maximal control invariant subset of the state constraints and that the implicitly defined safe set is stable. 
The concept of PCBF relates predictive safety filters to discrete-time control barrier functions (CBFs), such as proposed, e.g., in \cite{agrawal2017}. 
A main advantage of CBFs, see, e.g., \cite{ames2019} for an overview for continuous-time systems, is their ability to ensure stability of a safe subset of constraints as well as a dampening effect when approaching the boundary of the safe set. This is achieved by including a decrease constraint of an explicit CBF in an optimisation problem. Through its predictive nature, the PCBF algorithm in \cite{wabersich2022} enhances explicit CBFs by planning a trajectory towards their domain of attraction.
Another work which leverages the PCBF formulation with a CBF decrease constraint is given in \cite{didier2024approximate}, where an explicit approximation of the implicitly defined PCBF is used in a safety filter formulation. While this method provides input-to-state stability guarantees with respect to possible approximation errors, the method proposed in this work is guaranteed to converge to the exact safe sublevel set of the PCBF.
The PCBF method is extended in \cite{cortez2023robust} to a robust setting, however Lipschitz-based constraint tightenings, which grow as a function of the prediction horizon are used to ensure robust constraint satisfaction and closed-loop stability of a subset of the constraints is not considered. Compared to this formulation, the proposed RPI based tightening is independent of the prediction horizon and we provide rigorous convergence guarantees for the closed-loop system to a subset of the target region.
The PCBF algorithm can also be seen as a type of zone MPC, proposed for stabilisation of steady-states within a target set in \cite{gonzalez2009robust, ferramosca2010mpc} and generalised in \cite{liu2019model} to the stabilisation of control invariant sets. By leveraging a PCBF decrease constraint, the proposed formulation allows for optimising over a primary objective in a single optimisation problem compared to these approaches. 
Finally, Lyapunov decrease constraints for setpoint stabilisation have been proposed using explicit Lyapunov functions in, e.g., \cite{Mhaskar2006} and \cite{Heidarinejad2012}, and in a suboptimal or multiobjective approach in \cite{Scokaert1999,Pannocchia2011,Zeilinger2014,allan2017inherent,Mcallister2023}. Compared to these works, we propose a formulation which robustly converges to a desired target set rather than a desired setpoint by design of the optimisation problem.

\textit{Notation.} 
The distance of a vector $a$ to a set $\mathcal{A}$ is given by $\Vert x\Vert_\mathcal{A}:=\inf_{y\in\mathcal{A}} \Vert x-y\Vert$. We denote a sequence of vectors $x_{i|k}\in\mathbb{R}^n$ as $\bm{x}_k\coloneqq\{x_{i|k}\}_{i=0}^{N}$ where the sequence length $N$ can be deduced from context. The set $\mathcal{I}_{[a,b]}$ with $a,b\in\mathbb{N}$ denotes all integers $n\in\mathbb{N}$ such that $a\leq n \leq b$.

\section{Preliminaries and Problem Statement}\label{sec:prel}
We consider a continuous, discrete-time system 
\begin{equation}\label{eq:sys}
    x_{k+1}=f(x_k,u_k,w_k),
\end{equation}
where the state $x_k\in\mathbb{R}^n$, the input $u_k\in\mathbb{R}^m$ and the exogenous disturbance lies in a compact, known set $\mathcal{W}$ which contains the origin in its interior, i.e., $w_k\in\mathcal{W}\subset\mathbb{R}^n$, for all $k\geq 0$. The system is subject to compact input constraints, which can arise, e.g., from physical limitations of actuators of the form
$$u_k\in\mathcal{U} \;\forall k\geq 0.$$
The considered control objective is two-fold. Firstly, we consider a set of states which is deemed safe for operation of the system, i.e. the aim is to satisfy
$$x_k\in\mathcal{X}\coloneq\{x\in\mathbb{R}^n \mid c_x(x)\leq 0 \},$$
where $c_x:\mathbb{R}^n\rightarrow \mathbb{R}^{n_x}$ is continuous.
Note that it does not necessarily hold that the system state $x_0$ is originally contained within this set and therefore a soft formulation of the constraints is considered, i.e., 
$$\mathcal{X}(\xi)\coloneq\{x\in\mathbb{R}^n\mid c_x(x)\leq \xi\},$$
where $\xi\in\mathbb{R}^{n_x}$ is a slack variable, which is a formulation typically employed in practical predictive optimisation based applications, see, e.g., \cite{kerrigan2000soft}. 
The desired system behaviour can be characterised as follows. The system state should remain within a control invariant subset of $\mathcal{X}$. If the system state lies outside $\mathcal{X}$ due to initialisation of the system state or unexpected disturbances $w_k\notin\mathcal{W}$, it should converge back to the set for all possible disturbances $\mathcal{W}$.
Such a requirement could, e.g., arise from a lane changing manoeuvre on a highway or reaching a safe range of temperatures in a chemical reactor. 

Secondly, there is a primary control objective which should be minimised, i.e., a cost $J(\bm{x}_k,\bm{u}_k,p_k)$ for some time-varying parameter $p_k$ which might only be known at time step $k$. The primary control objective can be of economic nature, i.e. minimising cost of operation, a comfort metric during a lane changing manoeuvre, or a safety filter cost $\Vert u_k - p_k \Vert$, where $p_k$ is a control input proposed from a human operator or learning-based controller, see, e.g., \cite{wabersich2023} for an overview of safety filters and their applications. 

Finally, for ease of readability, we restate the assumptions on the system dynamics and the constraints in this section as follows.
\begin{assumption}\label{ass:prel}
    The system dynamics $f:\mathbb{R}^n\times\mathbb{R}^m\times\mathbb{R}^n\rightarrow\mathbb{R}^n$ are continuous and the set $\mathcal{W}\subset\mathbb{R}^n$ is compact and contains the origin in its interior. The input constraints $\mathcal{U}$ are compact, the function defining the state constraints $c_x$ is continuous and $\mathcal{X}(\xi)$ is compact for all $0\leq\xi<\infty$.
\end{assumption}

In the following, we summarise the nominal PCBF framework, which relies on the solution of two optimisation problems, in Section~\ref{sec:PCBF} and propose a robust extension in Section~\ref{sec:rPCBF} before proposing a single multiobjective optimisation problem in Section~\ref{sec:MOPCBF}. The approach is validated in numerical examples in Section~\ref{sec:NE}.

\section{Preliminaries on Predictive Control Barrier Functions} \label{sec:PCBF}
In this section, we review the concept of predictive control barrier functions (PCBFs), which was originally proposed in \cite{wabersich2022} in order to provide stability guarantees of a control invariant subset $\mathcal{S}\subseteq\mathcal{X}$ within some domain of attraction $\mathcal{D}\supset\mathcal{S}$, while minimising a primary safety filter objective. The algorithm proposed in \cite{wabersich2022} considers nominal dynamics of the form
\begin{equation}\label{eq:sysnom}
    z_k = f(z_k, v_k, 0),
\end{equation}
where $z_k\in\mathbb{R}^n$ is a nominal state and $v_k\in\mathbb{R}^m$ a nominal input, implying perfect knowledge of the underlying dynamics of the considered system. 

In order to achieve the desired system behaviour, the notion of control barrier functions (CBFs) is used, which enable establishing invariance of a set $\mathcal{S}$ and its stability within some domain of attraction $\mathcal{D}$.

\begin{definition}[Control Barrier Function \cite{wabersich2022}]~\\
\label{def:CBFcont}
A function $h: \mathcal{D} \to \mathbb{R}$ is a nominal discrete-time control barrier function with safe set $\mathcal{S} = \{z \in \mathbb{R}^n \mid h(z) \leq 0 \} \subset \mathcal{D}$
 if the following hold:\begin{enumerate}
    \item[(1)] $\mathcal{S}, \mathcal{D}  $ are compact and non-empty sets,
    \item[(2)] $h(z)$ is continuous in $\mathcal{D}$,
    \item[(3)] $\exists \Delta h : \mathcal{D} \to \mathbb{R}$, with $\Delta h$ continuous and $\Delta h(z)>0 $ for all $z \in \mathcal{D} \setminus \mathcal{S} $ such that:
\begin{subequations}
\label{eq:rate_cond_cbf}
\begin{align}
    &\forall z \in \mathcal{D} \setminus \mathcal{S} :\label{eq:outside_cbf_condition} \\ & \inf_{v\in \mathcal{U}} \{h(f(z,v,0))\mid f(z,v,0)\in\mathcal{D}\}- h(z) \leq - \Delta h(z) , \nonumber \\
    &\forall z \in \mathcal{S} \: : \inf_{v\in \mathcal{U}}h(f(z,v,0)) \leq 0. \label{eq:inside_cbf_condition}
\end{align}
\end{subequations}
\end{enumerate}
\end{definition}

The aim in \cite{wabersich2022} is to use a predictive control method to enhance an explicit, possibly conservative, CBF $h_f$ for a nominal system such that a primary objective can be optimised. This is achieved by showing that its proposed control strategy implies the existence of a CBF, such that a set $\mathcal{S}\supseteq\mathcal{S}_f\coloneq\{z\in\mathbb{R}^n\mid h_f(z)\leq 0\}$ is invariant and asymptotically stable within a larger domain $\mathcal{D}\supseteq\mathcal{D}_f\coloneq\{z\in\mathbb{R}^n\mid h_f(z)\leq \gamma_f\}$ for some $\gamma_f\in\mathbb{R}_{>0}$. 
The optimisation problem proposed in \cite{wabersich2022} which is solved at every time step is then given by
\begin{subequations}
\label{eq:pcbf_sf}
\begin{align}
 \min_{\bm{z}_k, \bm{v}_{k}} \; \; & J(\bm{z}_{k}, \bm{v}_{k}, p_k) \\
\text{s.t.}\; \; &\forall\, i = 0,\dots,N-1, \nonumber \\
& z_{0|k} = z_k, \label{eq:PCBF_init}\\
& z_{i+1|k} = f(z_{i|k}, v_{i|k}, 0),\label{eq:PCBF_dyn} \\
& v_{i|k} \in \mathcal{U}, \label{eq:PCBF_inp} \\
& z_{i|k} \in \overline{\mathcal{X}}_i( \xi^*_{i|k}), \label{eq:PCBF_state} \\
&h_f(z_{N|k}) \leq  \xi^*_{N|k}, \label{eq:PCBF_term}
\end{align}
\end{subequations}
where the input applied to the system at every time step is given by an optimal solution to the optimisation problem \eqref{eq:pcbf_sf}, i.e. $v_k=v^*_{0|k}$. Initialised at the current nominal state $z_k$ in \eqref{eq:PCBF_init}, a trajectory of states is predicted according to the nominal system dynamics \eqref{eq:PCBF_dyn}, such that the inputs which are optimised satisfy the required input constraints in \eqref{eq:PCBF_inp}. In order to establish stability guarantees, the state constraints enforced on the predicted trajectory in \eqref{eq:PCBF_state} are subject to an additional tightening, i.e., $\bar{\mathcal{X}}_i(\xi)\coloneqq\{z\in\mathbb{R}^n \mid c_x(z) \leq \xi - \Delta_i \mathbf{1}\}$, where it holds that $\Delta_{i+1}>\Delta_i>0=\Delta_0$ for all $i\in\{1,\dots,N-1\}$. Finally, the terminal constraint \eqref{eq:PCBF_term} is stated in terms of the function $h_f$, which is assumed to be a CBF according to Definition~\ref{def:CBFcont}.

\begin{assumption}\label{ass:termCBF}
 The function $h_f:\mathcal{D}_f \rightarrow \mathbb{R}$ with domain $\mathcal{D}_f:=\{z\in\mathbb{R}^n \mid h_f(z) \leq \gamma_f \},$ with $\gamma_f>0$, and corresponding safe set $\mathcal{S}_f:=\{z\in\mathbb{R}^n \mid h_f(z) \leq 0 \}\subseteq \overline{\mathcal{X}}_{N-1}(0)$ is a CBF according to Definition~\ref{def:CBFcont} for the nominal system dynamics \eqref{eq:sysnom} and is known. 
\end{assumption}

We note that the slack variables $\xi^*_{i|k}$ in the state constraints \eqref{eq:PCBF_state} and $\xi^*_{N|k}$ in the terminal constraint \eqref{eq:PCBF_term} are not optimisation variables in \eqref{eq:pcbf_sf}, but are precomputed as an optimal solution to another optimisation problem, which aims to minimise the total constraint violations over the prediction horizon:
\begin{subequations}
\label{eq:pcbf}
\begin{align}
 \hpb(z_k) := \min_{\bm{z}_k, \bm{v}_{k}, \bm{\xi}_{k}} \; \; & \alpha_f \xi_{N|k} + \sum_{i=0}^{N-1}\lvert \lvert{\xi_{i|k}}\rvert \rvert \label{eq:cbf_pcbf} \\
\text{s.t.}\; \; &\forall\, i = 0,\dots,N-1, \nonumber \\
& z_{0|k} = z_k, \label{eq:slack_init_pred} \\
& z_{i+1|k} = f(z_{i|k}, v_{i|k}, 0), \\
& v_{i|k} \in \mathcal{U}, \label{eq:slack_input_constraint}\\
& z_{i|k} \in \overline{\mathcal{X}}_i(\xi_{i|k}), \; 0 	\leq \xi_{i|k}, \label{eq:pcbf_state_constraint}\\
& h_f(z_{N|k}) \leq \xi_{N|k}, \; 0 \leq \xi_{N|k}.
\end{align}
\end{subequations}

Given this predictive control based slack computation, it is shown in \cite{wabersich2022}, that the optimal value function of \eqref{eq:pcbf}, i.e., $\hpb$, is a CBF according to Definition~\ref{def:CBFcont}. Therefore it holds that the set $\Spb\coloneqq\{z\in\mathbb{R}^n \mid \hpb(z)\leq 0 \}$ is stable in an enlarged domain $\Dpb\supseteq\mathcal{D}_f$. Furthermore, due to the fact that $\hpb(z_k)=0$ implies that the predicted nominal state trajectory does not violate tightened state constraints, it holds that $\Spb\subseteq \mathcal{X}$. The theoretical result is formalised in \cite[Theorem III.6]{wabersich2022}.


\section{Robust predictive control barrier function}\label{sec:rPCBF}
In this section, we introduce a robust version of the predictive control barrier algorithm illustrated in Section~\ref{sec:PCBF}. As for practical applications, a perfect system model is generally not available, 
the need arises to systematically consider persistent disturbances in the system model, as described in \eqref{eq:sys}, and control algorithm. The robust control approach which is proposed aims to enhance the existing nominal formulation of the PCBF algorithm in \cite{wabersich2022} with robustness guarantees while enabling the use of a nominal terminal CBF $h_f$ instead of a robust terminal CBF formulation. It relies on the design of a robust positive invariant (RPI) set with respect to the error between the true system state and a nominal trajectory. By ensuring that this error lies within the designed RPI set, robust convergence to a safe set implicitly defined through an optimisation problem can be guaranteed. 

\begin{assumption}\label{ass:RPI}
    A compact set $\mathcal{E}\subset\mathbb{R}^n$ and a continuous control law $\kappa:\mathbb{R}^n\rightarrow\mathbb{R}^m$ exist and are known, such that $\forall x, z\in\mathbb{R}^n$, 
    \begin{equation}
        x-z\in\mathcal{E} \Rightarrow f(x,v+\kappa(x-z), w) - f(z,v,0) \in \mathcal{E}.
    \end{equation}
\end{assumption}
An RPI set and corresponding error feedback controller according to Assumption~\ref{ass:RPI} can be designed, e.g., by leveraging incremental Lyapunov stability proposed for continuous-time systems in \cite{angeli2002lyapunov} and discrete-time systems in \cite{bayer2013discrete}. A possible synthesis method for discrete-time systems leverages semi-definite or sum-of-squares optimisation to optimise the control law and Lyapunov function for the differential dynamics of \eqref{eq:sys}, see, e.g. \cite{kohler2019nonlinear} for details. For linear system dynamics, the design simplifies to the design of a standard RPI set, through, e.g., polytopic methods in \cite{rakovic2005invariant} or the synthesis of quadratic Lyapunov functions in \cite{vanantwerp2000tutorial}.

The proposed robust PCBF scheme is based on tightening the constraints with respect to the RPI set $\mathcal{E}$, as originally proposed in \cite{mayne2005robust}. The approach leverages the fact that the error of a nominal state $z$ and with respect to the true state $x$ is contained within the RPI set $\mathcal{E}$ when applying the nominal inputs along with the error feedback $\kappa$ to the system. It therefore suffices to ensure that the nominal prediction in the PCBF optimisation problem satisfies the state and input constraints \eqref{eq:pcbf_state_constraint} and \eqref{eq:slack_input_constraint} tightened with the RPI set for the error $\mathcal{E}$ and the set of possible error feedbacks $\kappa(\mathcal{E})$, respectively. By using this constraint tightening, robust constraint satisfaction and convergence to a safe set can be established. As the constraints are additional tightened with the RPI set $\mathcal{E}$, the safe set $\mathcal{S}_f$ defined by the terminal control barrier function $h_f$ needs to be subset of $\bar{\mathcal{X}}_{N-1}(0)\ominus \mathcal{E}$ rather than $\bar{\mathcal{X}}_{N-1}(0)$.
\begin{assumption}\label{ass:rTermCBF}
    The terminal control barrier function $h_f$ satisfies Assumption~\ref{ass:termCBF} with respect to the tightened input set $\mathcal{U}\ominus\kappa(\mathcal{E})$.
    Additionally, it holds that $\mathcal{S}_f\coloneqq\{x\in\mathbb{R}^n \mid h_f(x)\leq 0\}\subseteq \bar{\mathcal{X}}_{N-1}(0)\ominus \mathcal{E}$. 
\end{assumption}

We note that by using the RPI set $\mathcal{E}$ for tightening the constraints, the same design procedure for the terminal CBF $h_f$ can be used as for the nominal system with the tightened state and input constraints. 
The optimisation problem to determine the optimal slack variables at every time step and its optimal value function are then given by
\begin{subequations}\label{eq:rPCBF}
\begin{align}
 h(x_k)\coloneqq\min_{\bm{z}_{k}, \bm{v}_{k}, \bm{\xi}_{k}} \; \; & \alpha_f \xi_{N|k} + \sum_{i=0}^{N-1}\lvert \lvert{\xi_{i|k}}\rvert \rvert \\
\text{s.t.}\; \; &\forall\, i = 0,\dots,N-1, \nonumber \\
&  x_k \in z_{0|k}\oplus\mathcal{E}, \label{eq:rPCBFinit} \\
& z_{i+1|k} = f(z_{i|k}, v_{i|k}, 0), \label{eq:rPCBFdyn} \\
& v_{i|k} \in \mathcal{U}\ominus \kappa(\mathcal{E}), \label{eq:rPCBFinp} \\
& z_{i|k} \in \overline{\mathcal{X}}_i( \xi_{i|k})\ominus \mathcal{E}, \;\xi_{i|k}\geq 0, \label{eq:rPCBFstate} \\
&h_f(z_{N|k}) \leq  \xi_{N|k}, \;\xi_{N|k}\geq 0. \label{eq:rPCBFterm}
\end{align}
\end{subequations}
Given an optimal solution for the slack variables $\bm{\xi}^*_k$ for \eqref{eq:rPCBF}, the primary objective can be optimised as follows:
\begin{subequations}\label{eq:rPCBFSF}
\begin{align}
 \min_{\bm{z}_{k}, \bm{v}_{k}} \; \; & J(\bm{z}_{k}, \bm{v}_{k}, p_k) \label{eq:rPCBFSFcost}\\
\text{s.t.}\; \; &\forall\, i = 0,\dots,N-1, \nonumber \\
&  x_k \in z_{0|k}\oplus\mathcal{E}, \\
& z_{i+1|k} = f(z_{i|k}, v_{i|k}, 0), \\
& v_{i|k} \in \mathcal{U}\ominus \kappa(\mathcal{E}), \\
& z_{i|k} \in \overline{\mathcal{X}}_i( \xi^*_{i|k})\ominus \mathcal{E},\\
&h_f(z_{N|k}) \leq  \xi^*_{N|k}.
\end{align}
\end{subequations}
After computing an optimal solution $v^*_{0|k}, z^*_{0|k}$ to the optimisation problem \eqref{eq:rPCBFSF}, the input applied to the system is given by $u_k=v^*_{0|k}+\kappa(x_k-z^*_{0|k})$, where we note that for a predictive safety filter application, the cost \eqref{eq:rPCBFSFcost} should minimise the difference between the proposed input $p_k$ and the input to be applied $u_k$. The theoretical guarantees of the corresponding control algorithm, detailed in Algorithm~\ref{alg:rpcbf} are then given as follows.

\begin{algorithm}[!t]
\caption{Robust PCBF}\label{alg:rpcbf}
\begin{algorithmic}[1]
\FOR{$k\geq 0$}
\STATE Measure state $x_k$
\STATE Obtain the parameter $p_k$
\STATE Compute optimal slack variables $\bm{\xi}^*_k$ \eqref{eq:rPCBF}
\STATE Compute an optimal solution of \eqref{eq:rPCBFSF} using $\bm{\xi}^*_k$
\STATE Apply $u_k\leftarrow v^*_{0|k}+\kappa(x_k-z^*_{0|k})$
\ENDFOR
\end{algorithmic}
\end{algorithm}

\begin{theorem}\label{thm:rPCBF}
    Let Assumptions~\ref{ass:prel},~\ref{ass:RPI}~and~\ref{ass:rTermCBF} hold. The optimisation problems \eqref{eq:rPCBF} and \eqref{eq:rPCBFSF} are recursively feasible and it holds that the set $\mathcal{S}\coloneqq\{x\in\mathbb{R}^n\mid h(x)\leq 0\}$ is robust asymptotically stable within $\mathcal{D}\coloneqq\{x\in\mathbb{R}^n \mid h(x)\leq \alpha_f \gamma_f\}$ for the system \eqref{eq:sys} under application of Algorithm~\ref{alg:rpcbf}.
\end{theorem}
\textit{Proof.}
We show recursive feasibility of the proposed algorithm as well as asymptotic stability of the set $\mathcal{S}\coloneqq\{x\in\mathbb{R}^n\mid h(x)\leq 0\}$ for the closed-loop system as follows. We show continuity and positive definiteness of $h$ despite the added, hard constraint on the initial state \eqref{eq:rPCBFinit}, as well as compactness of $\mathcal{S}$ and the domain of attraction of $h$, i.e. $\mathcal{D}\coloneqq\{x\in\mathbb{R}^n \mid h(x)\leq \alpha_f \gamma_f\}$. Next, we propose a candidate solution given a previously feasible solution to the optimisation problem \eqref{eq:rPCBF} which is also feasible for \eqref{eq:rPCBFSF} and establish that $h$ is a CBF according to Definition~\ref{def:CBFcont} 
through an appropriate decrease of $h$ at every time step and invariance of $\mathcal{S}$. 

\textit{1) Continuity of $h$ in $\mathcal{D}$.} Similarly to \cite[Theorem III.6]{wabersich2022}, we show continuity by showing that for every $\epsilon > 0$, there exists $\delta > 0$ such that $\Vert x - \bar{x} \Vert < \delta$ implies $|h(x)-h(\bar{x})|<\epsilon$ for any $x,\bar{x}\in\mathcal{D}$. Given an optimal solution to \eqref{eq:rPCBF} at $x$, we propose a suboptimal solution at $\bar{x}$ such that uniform continuity of the corresponding suboptimal value function $\bar{h}(\bar{x})$ can be established. Consider an optimal initial nominal state $z^*_0(x)$ and input sequence $\bm{v}^*(x)$. We construct a suboptimal solution at $\bar{x}$ by shifting the optimal nominal input $z^{*}_0(x)$ by $\bar{x}-x$, i.e. $\bar{z}_0^*(\bar{x})\coloneqq z^{*}_0(x) + \bar{x}-x$, which satisfies constraint \eqref{eq:rPCBFinit} by construction. The proposed suboptimal input sequence is equivalent to the optimal nominal input sequence at $x$, i.e. $\bar{\bm{v}}(\bar{x})\coloneqq \bm{v}^*(x)$. The state predictions $\bar{z}_{i}(\bar{x})$ given the suboptimal input sequence are then computed according to the nominal dynamics in \eqref{eq:rPCBFdyn} and the corresponding slacks can be computed as $\bar{\xi}_i(\bar{x})\coloneq \max(0, c_x(\bar{z}_i(\bar{x})) + \Delta_i \mathbf{1})$ and $\bar{\xi}_N(\bar{x})\coloneqq \max(0, h_f(\bar{z}_N(\bar{x})))$. The proposed suboptimal solution is therefore feasible, as $\bm{v}^*(x)$ and therefore $\bar{\bm{v}}^*(\bar{x})$ satisfy the input constraints \eqref{eq:rPCBFinp} and the state and terminal constraint \eqref{eq:rPCBFstate} and \eqref{eq:rPCBFterm} are satisfied by definition of the slack variables.

It holds the proposed suboptimal solution for the states $\bar{z}_i(\bar{x})$ is continuous in $\bar{x}$ due to the composition of the continuous function $f$ and the corresponding slacks $\bar{\xi}_i(\bar{x})$ are continuous due to the composition of the continuous functions $c_x$, or $h_f$ in the case of $\bar{\xi}_N$, $\bar{z}_i$ and the $\max$ operation. According to \cite[Lemma C.3]{wabersich2022}, for all $\bar{x}\in\mathcal{D}$, the state trajectory $\bar{z}_i(\bar{x})$ will lie in a compact set, implying uniform continuity of $\bar{z}_i(\bar{x})$ and $\bar{\xi}_i(\bar{x})$ according to the Heine-Cantor theorem in \cite[Theorem 4.19]{rudin1964principles}. As the suboptimal value function $\bar{h}(\bar{x})$ is a composition of continuous functions, it is uniformly continuous for all $\bar{x}\in\mathcal{D}$. Therefore, we have existence of $\delta>0$, such that $\Vert x-\bar{x}\Vert < \delta$ implies $h(\bar{x})- h(x) \leq \bar{h}(\bar{x}) - h(x) < \epsilon$ for a given $\epsilon > 0$. Additionally, following the same arguments, we can apply the optimal input sequence and initial nominal state from $\bar{x}$ to $x$. From uniform continuity of the suboptimal solution, it holds that there exists $\delta > 0$, such that $\Vert x-\bar{x}\Vert < \delta$ implies $h(x)-h(\bar{x})\leq \bar{h}(x) - h(\bar{x})< \epsilon$, for any given $\epsilon>0$, which implies continuity of $h$. Finally, it holds that $h$ is positive definite in $\mathcal{D}$ with respect to $\mathcal{S}$ and that $\mathcal{S}$ and $\mathcal{D}$ are compact, which follows directly from \cite[Theorem III.6]{wabersich2022}.

\textit{2) Recursive feasibility.}
Due to the use of the RPI set $\mathcal{E}$ in the robust formulation of the PCBF problems, the candidate sequence used for showing recursive feasibility of the optimisation problems \eqref{eq:rPCBF} and \eqref{eq:rPCBFSF} is the same as in the nominal case. First, note that any optimal solution to \eqref{eq:rPCBF} is trivially feasible in \eqref{eq:rPCBFSF} as the constraints of both optimisation problems are the same. Then, given an optimal solution to \eqref{eq:rPCBFSF} at $x_k$, we can construct a feasible solution to \eqref{eq:rPCBF} in the next time step as follows
\begin{align}\label{eq:candseq}
    \tilde{\bm{v}}_{k+1} & = \left(v^*_{1|k}, v_{2|k}^*, \dots, v_{N-1|k}^*, v_k^*(z_{N|k}^*)\right), \nonumber \\
    \tilde{\bm{z}}_{k+1} & = \left(z^*_{1|k}, z^*_{2|k}, \dots, z_{N|k}^*, f(z_{N|k}^*,v_k^*(z_{N|k}^*), 0 )\right), \\
    \tilde{\xi}_{i|k+1}& =\begin{cases}
        \max(0,\xi^*_{i|k}+(\Delta_{i}-\Delta_{i+1})\mathbf{1}),\; i \in  \mathcal{I}_{[0,{N-2}]}, \\ 
        \max(0, c_x(z^*_{N|k})+\Delta_{N-1}\mathbf{1}), \; i = N-1, \\
        \max(0, h_f(f(z^*_{N|k},v^*_k(z^*_{N|k}),0))), \; i = N, \nonumber
    \end{cases}
\end{align}
where we dropped the dependencies on the state $x_k$ for ease of readability. The final input $v_k^*(z^*_{N|k})$ is a solution to \eqref{eq:rate_cond_cbf} and exists according to Assumption~\ref{ass:rTermCBF} as it holds that $z_{N|k}^*\in\mathcal{D}_f$ due to the definition of the domain $\mathcal{D}=\{x\in\mathbb{R}^n\mid h(x)\leq \alpha_f\gamma_f\}$. 
As the error $x_k-z^*_{0|k}$ is contained in the RPI set $\mathcal{E}$ by constraint \eqref{eq:rPCBFinit}, it holds that under application of $u_k=v^*_{0|k}+\kappa(x_k-z^*_{0|k})$, the error $x_{k+1}-z^*_{1|k}$ is also contained within $\mathcal{E}$, satisfying \eqref{eq:rPCBFinit} for the proposed candidate solution. Furthermore, it holds that $\tilde{v}_{i|k}\in\mathcal{U}\ominus \kappa(\mathcal{E})$ due to the use of the shifted optimal solution from the previous time step and the fact that the final input $v^*_k(x)\in\mathcal{U}\ominus \kappa(\mathcal{E})$ according to Assumption~\ref{ass:rTermCBF}. 
Finally, the nominal states $\tilde{z}_{i|k}$ follow the dynamics \eqref{eq:rPCBFdyn} and the slacked state and terminal constraints \eqref{eq:rPCBFstate} and \eqref{eq:rPCBFterm}, respectively, are satisfied by construction of $\tilde{\bm{\xi}}_k$ through the definition of the state constraints $\bar{\mathcal{X}}_i(\tilde{\xi}_{i|k})$ and the terminal constraint. It therefore holds that Algorithm~\ref{alg:rpcbf} consists of recursively feasible optimisation problems.

\textit{3) Asymptotic stability of $\mathcal{S}$.} In order to show asymptotic stability of $\mathcal{S}$ within $\mathcal{D}$, it suffices to show that for all $x(0)\in\mathcal{D}$,  $\lim_{k\rightarrow \infty} h(x_k) = 0$ due to the definition of $\mathcal{S}$ and that $\forall \epsilon > 0, \exists \delta > 0$ such that 
$\Vert x(0) \Vert_\mathcal{S} < \delta \Rightarrow \forall k > 0: \Vert x(k) \Vert_\mathcal{S} < \epsilon$. Given the feasible candidate sequence which was proposed in \eqref{eq:candseq}, the optimal value function $h$ incurs a decrease at every time step, which is given by
\begin{align*}
& \tilde{h}(\bm{\xi}_{k}^*(x_k))-\tilde{h}(\bm{\tilde{\xi}}_{k+1}(x_k)), \\
    = & \left( \alpha_f(\xi^*_{N|k}(x_k)-\tilde{\xi}_{N|k+1}(x_k)) - \Vert \tilde{\xi}_{N-1|k+1}(x_k) \Vert\right)  \\ 
    & + \Vert \xi^*_{0|k}(x_k) \Vert +\sum_{i=0}^{N-2} \Vert \xi^*_{i+1|k}(x_k) \Vert - \Vert\tilde{\xi}_{i|k+1}(x_k) \Vert \\
    \coloneqq & \Delta h (x_k),
\end{align*}
where we define $\tilde{h}(\bm{\xi}_k)=\alpha_f\xi_{N|k}+\sum_{i=0}^{N-1}\Vert \xi_{i|k}\Vert$.
It follows directly from \cite[Theorem III.6]{wabersich2022} that $\Delta h(x_k)$ is continuous and strictly positive for all $x_k\in\mathcal{D}\setminus\mathcal{S}$ and $\Delta h = 0$ for all $x\in\mathcal{S}$. Due to suboptimality of the proposed feasible solution, we therefore have
\begin{align*} 
h(x_{k+1}) \leq & \tilde{h}(\tilde{\bm{\xi}}_{k+1}(x_k)) \\
\leq & \tilde{h}(\bm{\xi}_k^*(x_k)) - \Delta h(x_k) \\
= & h(x_k) - \Delta h(x_k)
\end{align*}
Given that $\Delta h(x)>0$ for all $x\in\mathcal{D}\setminus\mathcal{S}$, it holds that the sequence $h(x_k)$ is decreasing and therefore $\lim_{k\rightarrow \infty }h(x_k)=\alpha$ for some $0\leq \alpha \leq \alpha_f \gamma_f$ as $0\leq h(x_k)\leq \alpha_f \gamma_f $ for all $x_k\in\mathcal{D}$. 

We can then show by contradiction that $\alpha=0$. Assume that $\alpha > 0$. It then holds that there exists a set $\mathcal{A}\coloneq \{x \mid h(x)<\alpha\}$, such that $x_k\notin \mathcal{A} \; \forall k\geq 0$. Furthermore, due to continuity of $h$, there exists an open set $\mathcal{B}\coloneq \{x \mid \Vert x \Vert_{\mathcal{S}} < \beta\}\subset \mathcal{A}$ for some $\beta > 0$. We can then define the minimum decrease of $h$ as 
\begin{align*}
    \gamma \coloneqq \min_{x\in\mathcal{D}\setminus \mathcal{B}} \Delta h(x).
\end{align*}
Note that this minimum exists due to continuity of $\Delta h$ and compactness of $\mathcal{D}\setminus \mathcal{B}$ as $\mathcal{B}$ is open. As $\Delta h>0$ for all $x\in\mathcal{D}\setminus \mathcal{S}$, it holds that $\gamma > 0$. We can therefore consider 
\begin{align*}
& h(x_k) \\
= & h(x_{k-1}) + h(x_k) - h(x_{k-1}) \\
= & h(x_0) + \sum_{i=0}^{k-1} h(x_{i+1}) - h(x_i) \\
\leq & h(x_0) + \sum_{i=0}^{k-1} - \Delta h(x_k) \\
\leq & h(x_0) - k\gamma,
\end{align*}
where the first inequality holds through a telescopic sum. As $h(x_0)\leq \alpha_f \gamma_f$ by assumption, there exists some $\bar{k}\in\mathbb{N}$, such that for all $k\geq \bar{k}$, the non-increasing sequence $h(x_k) < \alpha$, which contradicts the assumption that $\lim_{k\rightarrow \infty} h(x_k)=\alpha > 0$. It therefore holds that $\alpha=0$ and $\lim_{k\rightarrow\infty}h(x_k)=0$, which implies that $\lim_{k\rightarrow \infty} x_k \in \mathcal{S}$ under application of Algorithm~\ref{alg:rpcbf}.
Finally, the fact that $\forall x(0)\in\mathcal{D}$, it holds that $\forall \epsilon > 0$, $\exists \delta > 0$ such that $\Vert x(0)\Vert_\mathcal{S} < \delta$ implies that $\forall k>0$, $\Vert x(k)\Vert_\mathcal{S} < \epsilon$ follows directly from \cite[Theorem A.3]{wabersich2022}. Consider $h^{-}\coloneqq \min_{\Vert x \Vert_\mathcal{S}\geq \epsilon, x\in\mathcal{D}} h(x)$, which exists due to continuity of $h$ and compactness of the domain of optimisation. By selecting $\delta > 0$ such that $\Vert x-\bar{x}\Vert <\delta \Rightarrow \Vert h(x)-h(\bar{x})\Vert < h^{-}$, such that $\Vert x(0)\Vert_\mathcal{S}<\delta$ implies $h(x(0))<h^{-}$, it holds through the fact that $h(x(k))$ is non-increasing that $h(x(k)) < h^{-} \leq h(x)$ for all $x$ where $\Vert x \Vert_\mathcal{S} \geq \epsilon$. Therefore, the set $\mathcal{S}$ is robustly asymptotically stable within $\mathcal{D}$.
$\hfill \qed$

\section{Multiobjective Predictive Control Barrier Function} \label{sec:MOPCBF}
In this section, we propose a multiobjective optimisation approach to solve the robust PCBF problem. While the proposed robust PCBF method in Section~\ref{sec:rPCBF} satisfies the requirements of the safe set $\mathcal{S}\subseteq\mathcal{X}$ being asymptotically stable within $\mathcal{D}$ and ensuring its forward invariance, it suffers from two main drawbacks which are inherited from the nominal PCBF algorithm in \cite{wabersich2022}. Firstly, the algorithm requires to solve two consecutive optimisation problems to determine the input $u_k$ to be applied to the system. Secondly, by minimising the constraint violations in \eqref{eq:rPCBF} at every time step, the feasible set of inputs to be applied to the system when the primary objective is optimised in \eqref{eq:rPCBFSF} is significantly reduced. Therefore, we propose an approach based on multiobjective and suboptimal model predictive control approaches, such as, e.g. in \cite{soloperto2020augmenting, allan2017inherent}. 
By introducing a CBF decrease constraint in the minimisation of the primary objective, the same stability guarantees can be recovered as for the two step approach in Section~\ref{sec:rPCBF} with a single optimisation problem. An additional benefit lies in the fact that we do not require the minimal constraint violations to be achieved at every time step, but rather a decrease with respect to a warmstart value for the PCBF, such that the primary objective can be more effectively minimised.

In order to impose a decrease constraint in a suboptimal model predictive control type fashion, we leverage the fact that a feasible solution of the predictive control barrier function problem \eqref{eq:rPCBF} can be constructed, such that a cost decrease is ensured. By imposing that the CBF value 
lies below the value of the CBF value of the known feasible, denoted warmstart, solution, it is ensured that the CBF decreases by at least the decrease of the warmstart solution. Due to the fact that the feasible solution in the proof of Theorem~\ref{thm:rPCBF}, specifically the last applied input $v^*_k(z^*_{N|k})$ is known to exist but not explicitly available, we assume that during the design of the nominal, terminal CBF $h_f$, a control law $\pi_f$ is additionally designed which satisfies \eqref{eq:rate_cond_cbf}, which is typical in the design of invariant sets and corresponding CBFs discussed in Section~\ref{sec:rPCBF}.

\begin{assumption}\label{ass:termPolicy}
    There exists a continuous control law $\pi_f:\mathbb{R}^n\rightarrow \mathbb{R}^m$ which is known such that 
    \begin{align*}
        \forall z\in\mathcal{D}_f\setminus\mathcal{S}_f &: \; h(f(z,\pi_f(z),0))-h(z)\leq -\Delta h_f(z) \\
        \forall z\in\mathcal{S}_f&: \; h(f(z,\pi_f(z),0))\leq 0
    \end{align*}
\end{assumption}
Given this terminal control law, we can define the explicit warmstart solution at time step $k+1$ given any sequence $(\bm{z}_k, \bm{v}_k, \bm{\xi}_k)$, as follows
\begin{subequations}\label{eq:warmstart}
    \begin{align}
    &\tilde{\bm{v}}_{k+1}(\bm{z}_k, \bm{v}_k, \bm{\xi}_k)  {=} \left(v_{1|k}, \dots, v_{N-1|k}, \pi_f(z_{N|k})\right)\!,  \\
    &\tilde{\bm{z}}_{k+1}(\bm{z}_k, \bm{v}_k, \bm{\xi}_k)  {=} \left(z_{1|k}, \dots, z_{N|k}, f(z_{N|k},\pi_f(z_{N|k}), 0 )\right)\!,  \\
    &\tilde{\xi}_{i|k+1}(\bm{z}_k, \bm{v}_k, \bm{\xi}_k) \nonumber \\
    =&\begin{cases}
        \max(0,\xi_{i|k}+(\Delta_{i}-\Delta_{i+1})\mathbf{1}),\; i \in  \mathcal{I}_{[0,{N-2}]}, \\ 
        \max(0, c_x(z_{N|k})+\Delta_{N-1}\mathbf{1}), \; i = N-1, \\
        \max(0, h_f(f(z_{N|k},\pi_f(z_{N|k}),0))), \; i = N.  \label{eq:slackwarmstart}
    \end{cases}
    \end{align}
\end{subequations}
We note that the warmstart sequence is only a function of the nominal state, input and slack sequences and does not depend on the disturbance which affects the true system dynamics, as is the case in \cite{didier2024predictive}. In fact, the warmstart sequence is the same as the feasible solution of the nominal PCBF problem \eqref{eq:pcbf} and the proposed formalism can therefore be used identically in a nominal formulation. This warmstart sequence causes a decrease 
\begin{align*}
    & \tilde{h}(\bm{\xi}_{k})-\tilde{h}(\bm{\tilde{\xi}}_{k+1}), \\
    = & \left( \alpha_f(\xi_{N|k}-\tilde{\xi}_{N|k+1}) - \Vert \tilde{\xi}_{N-1|k+1} \Vert\right)  \\
    & + \Vert \xi_{0|k} \Vert +\sum_{i=0}^{N-2} \Vert \xi_{i+1|k} \Vert - \Vert\tilde{\xi}_{i|k+1} \Vert \\
    \coloneqq& \Delta \tilde{h}\left(\bm{\xi}_{k}, \bm{\tilde{\xi}}_{k+1} \right), \\
\end{align*}
where we define 
\begin{align*}
\tilde{h}(\bm{\xi}_k) = \alpha_f\xi_{N|k} + \sum_{i=0}^{N-2} \Vert \xi_{i|k} \Vert. 
\end{align*}
It follows directly from the proof of Theorem~\ref{thm:rPCBF}, that this decrease is positive definite with respect to $\bm{\xi}_k\neq 0$, which will be leveraged to establish convergence guarantees of the proposed formulation.
The optimisation problem which is solved at every time step is then given by
\begin{subequations}\label{eq:robustPCBF}
\begin{align}
 \min_{\bm{z}_{k}, \bm{v}_{k}, \bm{\xi}_{k}} \; \; & J(\bm{z}_{k}, \bm{v}_{k}, p_k) \\
\text{s.t.}\; \; &\forall\, i = 0,\dots,N-1, \nonumber \\
&  x_k \in z_{0|k}\oplus\mathcal{E}, \\
& z_{i+1|k} = f(z_{i|k}, v_{i|k}, 0), \\
& v_{i|k} \in \mathcal{U}\ominus \kappa(\mathcal{E}), \\
& z_{i|k} \in \overline{\mathcal{X}}_i( \xi_{i|k})\ominus \mathcal{E}, \;\xi_{i|k}\geq 0,\\
&h_f(z_{N|k}) \leq  \xi_{N|k}, \;\xi_{N|k}\geq 0, \\
& \alpha_f \xi_{N|k} + \sum_{i=0}^{N-1}\Vert \xi_{i|k} \Vert \leq \nonumber \\ & \tilde{h}(\bm{\tilde{\xi}}_{k}) + c_\alpha \Delta \tilde{h}(\bm{\xi}^*_{k-1}, \bm{\tilde{\xi}}_k) \label{eq:rPCBFconstr}
\end{align}
\end{subequations}

The optimisation problem differs from \eqref{eq:rPCBFSF} only through the additional constraint \eqref{eq:rPCBFconstr}, which imposes that the PCBF value of the slacks which are optimised is bounded by the warmstart value. The additional term $c_\alpha\Delta \tilde{h}(\bm{\xi}^*_{k-1}, \bm{\tilde{\xi}}_k)$ allows reducing the required decrease of the PCBF value, with the hyperparameter $c_\alpha\in[0,1)$. Although the constraint \eqref{eq:rPCBFconstr} is formulated in terms of the slack variables $\xi_{i|k}$ and parameters $\tilde{\xi}_{i|k}$, the theoretical analysis for convergence of the closed-loop system is performed over the nominal state and warmstart inputs, as we have that $\tilde{\xi}_k$ is fully determined by $\tilde{z}_{0|k}$ and $\tilde{\bm{v}}_k$. We therefore define an augmented system of the nominal state $z^*_{0|k}$ and nominal input sequence $\bm{v}^*_k$ chosen by an optimiser as a feasible solution of \eqref{eq:robustPCBF} at time step $k$ as
\begin{equation*}
    s_k = (z^*_{0|k}, \bm{v}^*_k).
\end{equation*}
The considered augmented system $s_k$ evolves according to the following difference inclusion
\begin{alignat}{3}
    s_{k+1}  \in & \mathcal{H}(s_k, w_k)  &&  \\
     :=&\{(z^*_{0|k+1}, \bm{v}^*_{k+1}) {\mid} \; && (z^*_{0|k+1}, \bm{v}^*_{k+1})\! \in \! \mathcal{F}(x_{k+1}, \bm{\tilde{\xi}}_{k+1}, \bm{\xi}^*_k) \nonumber \\
    & && x_{k+1} {=} f(x_k, v^*_{0|k}{+}\kappa(x_{k}, z^*_{0|k}), w_k) \nonumber \\ 
    & && \bm{\tilde{\xi}}_{k+1} = \zeta(z^*_{0|k}, \bm{v}_k^*)\nonumber
    \},
\end{alignat}
where $\zeta$ denotes the computation of the sequence of warmstart slacks according to \eqref{eq:slackwarmstart} and $\mathcal{F}(x_k,\bm{\tilde{\xi}}_k, \bm{\xi}^*_{k-1})$ is the feasible set of \eqref{eq:robustPCBF} with parameters $x_k$, $\bm{\tilde{\xi}}_k$ and $\bm{\xi}^*_{k-1}$. The difference inclusion is initialised with $\bm{\tilde{\xi}}_{0}=\bm{\xi}^*$, solution of \eqref{eq:rPCBF} and $\Delta \tilde{h}(\bm{\tilde{\xi}}_0, \bm{\xi}^*_{-1})=0$, as detailed in Algorithm~\ref{alg:online}.

\begin{algorithm}[!t]
\caption{Robust Multiobjective PCBF}\label{alg:online}
\begin{algorithmic}[1]
\STATE Measure $x_0$
\STATE Initialise $\bm{\tilde{\xi}}_0 \leftarrow \bm{\xi}^*$, solution of \eqref{eq:rPCBF}
\STATE Initialise $\bm{\xi}^*_{-1}\leftarrow \bm{\tilde{\xi}}_0$
\FOR{$k\geq 0$}
\STATE Obtain the parameter $p_k$
\STATE Compute an optimal solution of \eqref{eq:robustPCBF} 
\STATE Apply $u_k\leftarrow v^*_{0|k}+\kappa(x_k-z^*_{0|k})$
\STATE Compute warmstart slacks $\bm{\tilde{\xi}}_{k+1}$ according to \eqref{eq:slackwarmstart}
\STATE Measure state $x_{k+1}$
\ENDFOR
\end{algorithmic}
\end{algorithm}

\begin{theorem}
Let Assumptions~\ref{ass:prel},~\ref{ass:RPI},~\ref{ass:rTermCBF}~and~\ref{ass:termPolicy} hold. The optimisation problem \eqref{eq:robustPCBF} is recursively feasible and system \eqref{eq:sys} converges to the set $\mathcal{S}=\{x\in\mathbb{R}^n \mid h(x)\leq 0\}$ under application of Algorithm~\ref{alg:online} within $\mathcal{D}=\{x\in\mathbb{R}^n \mid h(x)\leq \alpha_f \gamma_f\}$ for all $c_\alpha\in[0,1)$. 
\end{theorem}
\textit{Proof.}
Recursive feasibility of \eqref{eq:robustPCBF} follows directly from Part 2) of the proof of Theorem~\ref{thm:rPCBF} and the fact that the the feasible candidate solution \eqref{eq:candseq} is equivalent to the warmstart sequence \eqref{eq:warmstart} by selecting $v^*_k(z^*_{N|k})=\pi_f(z^*_{N|k})$, implying that the constraint \eqref{eq:rPCBFconstr} is satisfied for all $c_\alpha\in[0,1)$ due to $\Delta\tilde{h}(\bm{\xi}^*_{k-1}, \tilde{\xi}_k)\geq 0$.

We show that under application of Algorithm~\ref{alg:online}, the slack variables selected by the optimiser converge to $0$, i.e.
$$\lim_{k\rightarrow\infty}\tilde{h}(\bm{\xi}^*_k(s_k))=0,$$
which implies that $\lim_{k\rightarrow\infty} x_k \in \mathcal{S}$. We have 
\begin{align*}
& \tilde{h}(\bm{\xi}^*_{k+1}(s_{k+1})) \equiv \tilde{h}(s_{k+1}) \\
\leq & \sup_{(z^*_{0|k+1},\bm{v}_{k+1})\in\mathcal{F}(x_{k+1},\bm{\tilde{\xi}}_{k+1}, \bm{\xi}^*_{k})} \tilde{h}(z^*_{0|k+1}, \bm{v}_{k+1}) \\
\leq & \tilde{h}(\bm{\tilde{\xi}}_{k+1}) + c_\alpha \Delta \tilde{h}(\bm{\xi}^*_{k}, \bm{\tilde{\xi}}_{k+1}) \quad \textup{(by constraint \eqref{eq:rPCBFconstr})} \\
\leq & \tilde{h}(\bm{\xi}^*_{k}) - (1-c_\alpha) \Delta \tilde{h}(\bm{\xi}^*_{k},\bm{\tilde{\xi}}_{k+1}) \\
\equiv & \tilde{h}(s_k) - (1-c_\alpha) \Delta \tilde{h}(s_k) 
\end{align*}
It follows from \cite[Theorem III.6]{wabersich2022} that $(1-c_\alpha)\Delta \tilde{h}(s_k)>0$ for all $c_\alpha\in[0,1)$ and $s_k$ such that $\tilde{h}(s_k)>0$.
It therefore holds that $\lim_{k\rightarrow \infty} \tilde{h}(s_k) = \alpha$ for some $0\leq \alpha\leq \alpha_f\gamma_f$, as $\tilde{h}(s_k)\geq 0$ for all $s_k$ and $\tilde{h}(s_k)$ is a decreasing sequence with respect to $k$. Following the same arguments of the proof of Theorem~\ref{thm:rPCBF}, the case of $\alpha > 0$ can be shown to lead to a contradiction, such that it holds that $\lim_{k\rightarrow \infty} \tilde{h}(s_k) =0$, which implies that $\lim_{k\rightarrow\infty} x_k \in \mathcal{S}$ under application of Algorithm~\ref{alg:online}. $\hfill \qed$
\begin{remark}
    Compared to other predictive control approaches with Lyapunov decrease constraints, such as, e.g. \cite{allan2017inherent, didier2024predictive}, we do not require two separate warmstart generating functions for $\bm{\tilde{\xi}}_k$. This follows from the continuity of the value function $\tilde{h}$ and the decrease function $\Delta \tilde{h}$ as well as the fact that we do not consider a stage cost decrease in the provided analysis but rather a decrease as a function of the full prediction. 
\end{remark}

\section{NUMERICAL EXAMPLE} \label{sec:NE}
\subsection{Linear Space Rendezvous}\label{sec:space}
We consider a linear space rendezvous example according to the Clohessy-Wiltshire-Hill dynamics model, see, e.g., \cite{starek2017fast}, of the form
$$x_{k+1}=Ax_k+Bu_k+w_k.$$
The dynamics are unstable with the state $x_k\in\mathbb{R}^6$ consisting of a relative position $p_{\{x,y,z\}}$ and velocity $v_{\{x,y,z\}}$ in arbitrary units to a desired setpoint on a target orbit and the input $u_k\in\mathbb{R}^3$ consisting of force-per-unit-mass which can be applied by thrusters. The aim is to stabilise a spacecraft to a neighbourhood around the desired setpoint while using as little fuel as possible, i.e. minimising $\sum_{k=0}^{T} \Vert u_k \Vert_1$ over the task horizon $T=400$. The desired area is described by the constraints $x_k\in\mathcal{X}=\{x\in\mathbb{R}^6 \mid | p_{\{x,y,z\}}| \leq 10, |v_{\{x,y,z\}}| \leq 20 \}$ and the input constraints are given by $u_k\in\mathcal{U}=\{u\in\mathbb{R}^3 \mid \Vert u \Vert_\infty \leq 20\}$, which can be represented as $A_x x \leq b_x$ and $A_u u \leq b_u$, respectively. The exogeneous disturbance acts on the velocity only with the magnitude of the corresponding entries being bounded by $0.5$. 

\begin{figure*}[h]
    \centering
    \includegraphics[width=0.99\linewidth]{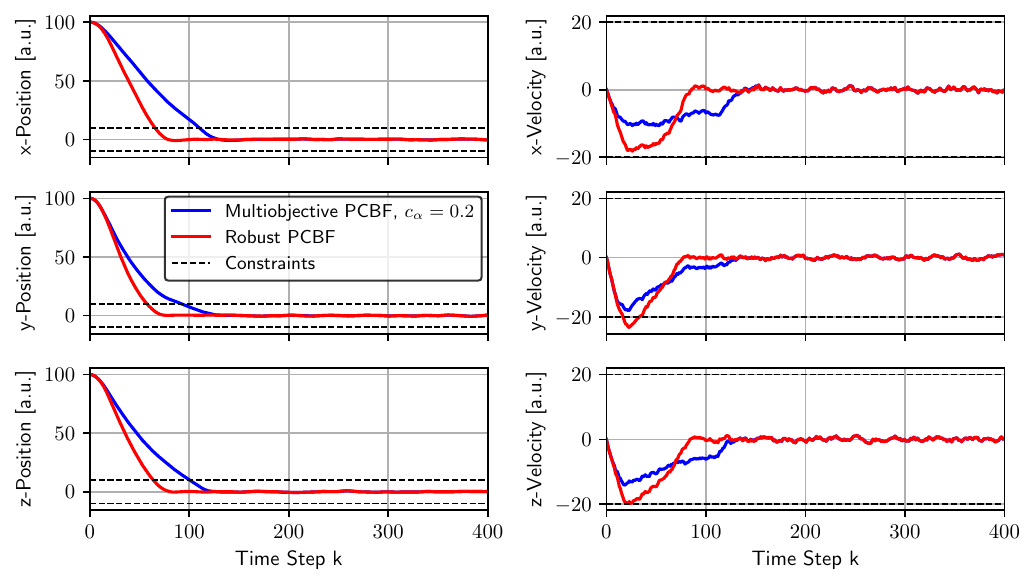}   
    \caption{Closed-loop state evolution for a space rendezvous example using the proposed robust PCBF formulation in Algorithm~\ref{alg:rpcbf} (\textit{red}) and the proposed multiobjective formulation in Algorithm~\ref{alg:online} (\textit{blue}). Using both approaches, the states are guaranteed to converge to the constraints (\textit{black}). An advantage of the multiobjective approach is to relax the required rate of convergence to the constraint set using the hyperparameter $c_\alpha\in[0,1)$.}
    \label{fig:lin_cl_state}
\end{figure*}

We design a terminal CBF $h_f(x)=x^\top P x - 1 $ and corresponding control law $\pi_f(x)=Kx$ according to the following semi-definite program (SDP), which maximises the volume of the corresponding ellipsoid, see, e.g. \cite{wabersich2018scalable}:
\begin{subequations}\label{eq:SDPCBF}
    \begin{align}
    \min_{E,Y} & -\log\det(E) \\
    \textup{s.t. } & E\succeq 0 \\
    &\begin{bmatrix} E & (AE+BY)^\top \\ AE+BY & E
    \end{bmatrix} \succeq 0 \\
    & \begin{bmatrix} ([\bar{b}_x]_i - \Delta_{N-1})^2 & [A_x]_iE \\ ([A_x]_iE)^\top & E 
    \end{bmatrix} \succeq 0 \;\; \forall i=1,\dots 4 \label{eq:SDPstateconstr}\\
    & \begin{bmatrix} [\bar{b}_u]_j^2 & [A_u]_jY \\ ([A_u]_jY)^\top & E 
    \end{bmatrix} \succeq 0 \;\; \forall j=1,\dots 2.\label{eq:SDPinputconstr}
    \end{align}
\end{subequations}
From this (SDP), we can recover $P=E^{-1}$ and $K=YP$. The tightened constraint values $\bar{b}_x$ and $\bar{b}_u$ in \eqref{eq:SDPstateconstr} and \eqref{eq:SDPinputconstr} are obtained by tightening the constraint set $\mathcal{X}$ and $\mathcal{U}$ with an RPI set $\mathcal{E}$ and the corresponding set of inputs $\kappa(\mathcal{E})$, respectively. This RPI set is designed in an SDP similar to \eqref{eq:SDPCBF}, which aims to minimise the constraint tightening with respect to $\mathcal{E}$. 

\begin{figure}
    \centering
    \includegraphics[width=0.99\linewidth]{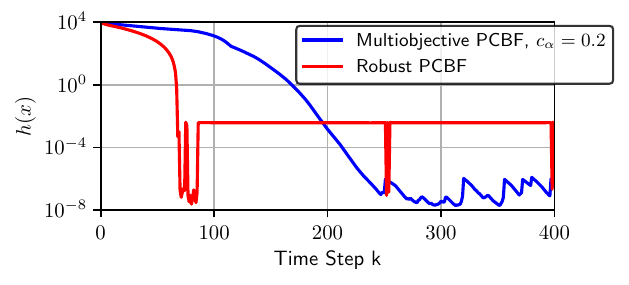}   
    \caption{Value of the predictive control barrier function for the robust approach in Algorithm~\ref{alg:rpcbf} (\textit{red}) and its upper bound for the multiobjective approach in Algorithm~\ref{alg:online} (\textit{blue}) The hyperparameter $c_\alpha$ allows tuning the required rate of decrease of $h$.}
    \label{fig:lin_cl_pcbf}
\end{figure}

Given the terminal CBF $h_f$ and RPI set $\mathcal{E}$, we compare the robust PCBF formulation in Algorithm~\ref{alg:rpcbf} and the proposed multiobjective approach in Algorithm~\ref{alg:online}, using $\Delta_i = 1$e-3 and $\alpha_f=1$e6, minimising the primary cost over the prediction horizon $N=200$, i.e. $J(\mathbf{z_k},\mathbf{v_k}, p_k) = \sum_{i=0}^N\Vert v_{i|k} \Vert_1$. The closed-loop state evolutions for an initial state with $p_{\{x,y,z\}}=100$ are given in Figure~\ref{fig:lin_cl_state} and the corresponding CBF values $h(x)$ in Figure~\ref{fig:lin_cl_pcbf}. Due to the theoretical guarantees established in Section~\ref{sec:rPCBF}~and~\ref{sec:MOPCBF}, the closed-loop is guaranteed to converge to the constraints $\mathcal{X}$ even in the presence of the persistent disturbance $w_k$. The hyperparameter $c_\alpha$ allows reducing the rate of convergence to the constraint set, which can be observed in the slower decrease of $h(x)$ in Figure~\ref{fig:lin_cl_pcbf}. This results in the benefit of a less conservative optimisation of the primary objective, i.e. the fuel consumption. The closed-loop cost $\sum_{i=0}^{400} \Vert u_k \Vert_1$ is given by $2038$ in the robust PCBF formulation and by $1551$ for the proposed multiobjective formulation, i.e. a decrease of $23.9\%$. We note that a major part of the fuel consumption is due to the disturbance compensation. When applying the algorithm without a disturbance present, the corresponding values are $1225$ and $809$, respectively, i.e. a decrease of $34\%$ using the multiobjective approach. Finally, the multiobjective approach provides another benefit in being computationally more efficient. The robust PCBF approach with two optimisation problems takes $234$ms per time step to compute using CVXPY \cite{diamond2016cvxpy} and MOSEK \cite{andersen2000mosek} on an Intel Core i7-10510U CPU @ 1.80GHz, 
while the multiobjective approach takes $148$ms, i.e. a decrease of $37\%$.

\subsection{Lane Changing Example}
We consider a nonlinear kinematic bicycle model using Euler discretisation, inspired by the miniature race car example in \cite{carron2023chronos} and which can be used for vehicles, of the form 
\begin{align*}
x_{k+1}&=f(x_k,u_k,w_k) \nonumber\\
&=\begin{bmatrix}
p_{y,k}+ T_s(v_k\sin(\psi_k) \\
\psi_k + T_s(v_k/l\tan(\delta_k)) + w_k \\
v_k + T_s \tau_k \\
\delta_k + \Delta \delta_k
\end{bmatrix}.
\end{align*}
The state is given by the lateral deviation to a lane $p_{y,k}$, the heading angle $\psi_k$, the velocity $v_k$ and the steering angle $\delta_k$, i.e. $x_k = [p_{y,k}, \psi_k, v_k, \delta_k ]\in\mathbb{R}^4$ and the input is given by the acceleration $\tau_k$ and the steering change $\Delta \delta_k$, i.e. $u_k = [\Delta \delta_k, \tau_k]$. The disturbance acts on the heading angle, with a magnitude of $0.5^\circ$ per time step $T_s=50$ms. The target lane width is given by $0.4$m and we impose additional requirements for the heading $|\psi_k|\leq 35^\circ$, the velocity $0.5$ms$^{-1}\leq v_k \leq 1.3$ms$^{-1}$ and the steering $|\delta_k|\leq 35^\circ$, resulting in the constraints $x_k\in\mathcal{X}=\{x\in\mathbb{R}^4 \mid c_x(x)\leq 0\}$. The input constraints are given by $|\Delta \delta_k|\leq 17.5^\circ$. 

The terminal CBF is computed by computing a Lyapunov function for the linearised system around the steady-state velocity $v_s=1.1$ms$^{-1}$ using an SDP and verifying invariance for the nonlinear dynamics according to \cite[Section IV]{wabersich2022}. The RPI design is similar to Section~\ref{sec:space}, where an SDP is solved for all possible linearisations of the dynamics for $x_k\in\mathcal{X}$ and all $p_y\in\mathbb{R}^n$ to synthesise an incremental Lyapunov function and corresponding linear feedback controller, such that robust invariance is guaranteed for the nonlinear system error for these states. 

\begin{figure}[h]
    \centering
    \includegraphics[width=0.99\linewidth]{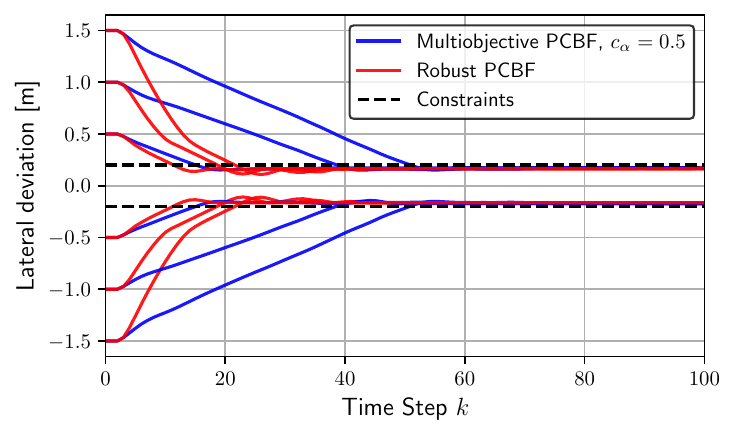}
    \caption{Closed-loop simulation of a human-operated vehicle with the proposed PCBF formulations, where the human driver aims to actively steer away from the desired lane.}
    \label{fig:nonlinear}
\end{figure}

The primary objective is to use a predictive safety filter formulation, which aims to match the inputs proposed by a human driver. The human driver is modelled as a linear control law, aiming to destabilise the car, i.e. steer the car away from the lane. The corresponding cost which is minimised online is therefore given as $J(\bm{z}_k,\bm{v}_k,p_k)=\Vert v_{0|k}+\kappa(x_k-z_{0|k}) - p_k\Vert$, where $p_k$ is the proposed human input at time $k$. The closed-loop convergence of the car to the lane with $N=20$ is demonstrated in Figure~\ref{fig:nonlinear}, where it can be observed that both formulations converge to the lane as desired. Using the multiobjective formulation with $c_\alpha=0.5$, a slower convergence is achieved. Finally, we note that after convergence, the vehicle closely follows the edge of the road even in the presence of disturbances. The computation times for the robust PCBF approach using CasADi \cite{andersson2019} and IPOPT \cite{biegler2009} are given by $140$ms per time step, compared to $116$ms for the multiobjective approach. 

\section{CONCLUSION} \label{sec:concl}
We propose a robust predictive control barrier function formulation, enabling convergence of a system state to a desired target set even in the presence of bounded, persistent disturbances. The formulation relies on a constraint tightening using a robust positive invariant set, ensuring recursive feasibility. The robust formulation consists of solving two optimisation problems which determine the minimal possible constraint violations in a first step and optimise a primary objective in a second step. By using a decrease constraint on the predictive control barrier function in a multiobjective fashion, we show that the formulation can be reduced to a single optimisation problem. The inclusion of a decrease constraint enables the introduction of a hyperparameter, which governs the rate of convergence to the desired target set. By requiring a less stringent convergence, a less conservative primary objective minimisation can be achieved. The proposed methods are validated on a linear space rendezvous example, demonstrating the reduction of conservativeness in a fuel optimisation setting as well as the computational benefits of the multiobjective formulation, and as a driver assistance system for a nonlinear kinematic bicycle model.

\bibliographystyle{plain}       
\bibliography{bibliography} 

\subsection*{  } 
\setlength\intextsep{0pt} 

\newpage
\begin{wrapfigure}{l}{0.13\textwidth}
    \centering
    \includegraphics[width=0.15\textwidth]{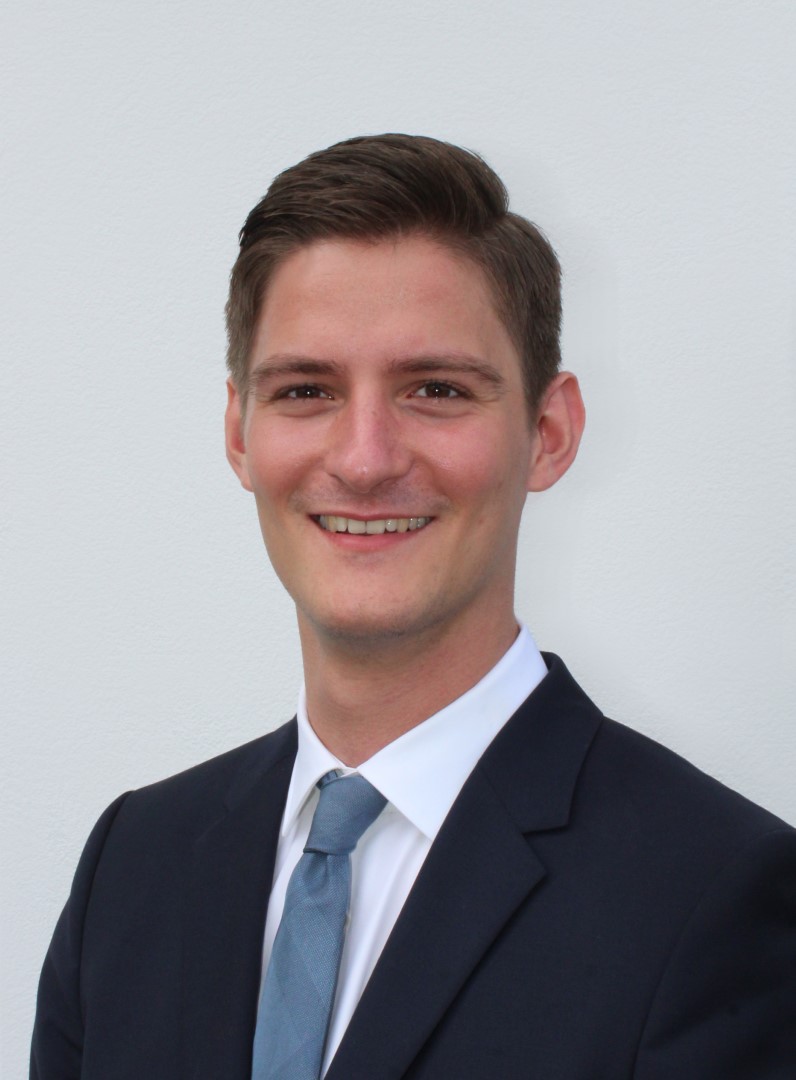}
\end{wrapfigure}
\noindent \textbf{Alexandre Didier} received his BSc. in Mechanical Engineering from ETH Zurich in 2018 and his MSc. in Robotics, Systems and Control from ETH Zurich in 2020. He is now pursuing a Ph.D. under the supervision of Prof. Dr. Melanie Zeilinger in the Intelligent Control Systems Group at the Institute of Dynamic Systems and Control (IDSC) at ETH Zurich. His research interests lie especially in model predictive control and regret minimisation for safety-critical and uncertain systems.

\begin{wrapfigure}{l}{0.13\textwidth}
    \centering
    \includegraphics[width=0.15\textwidth]{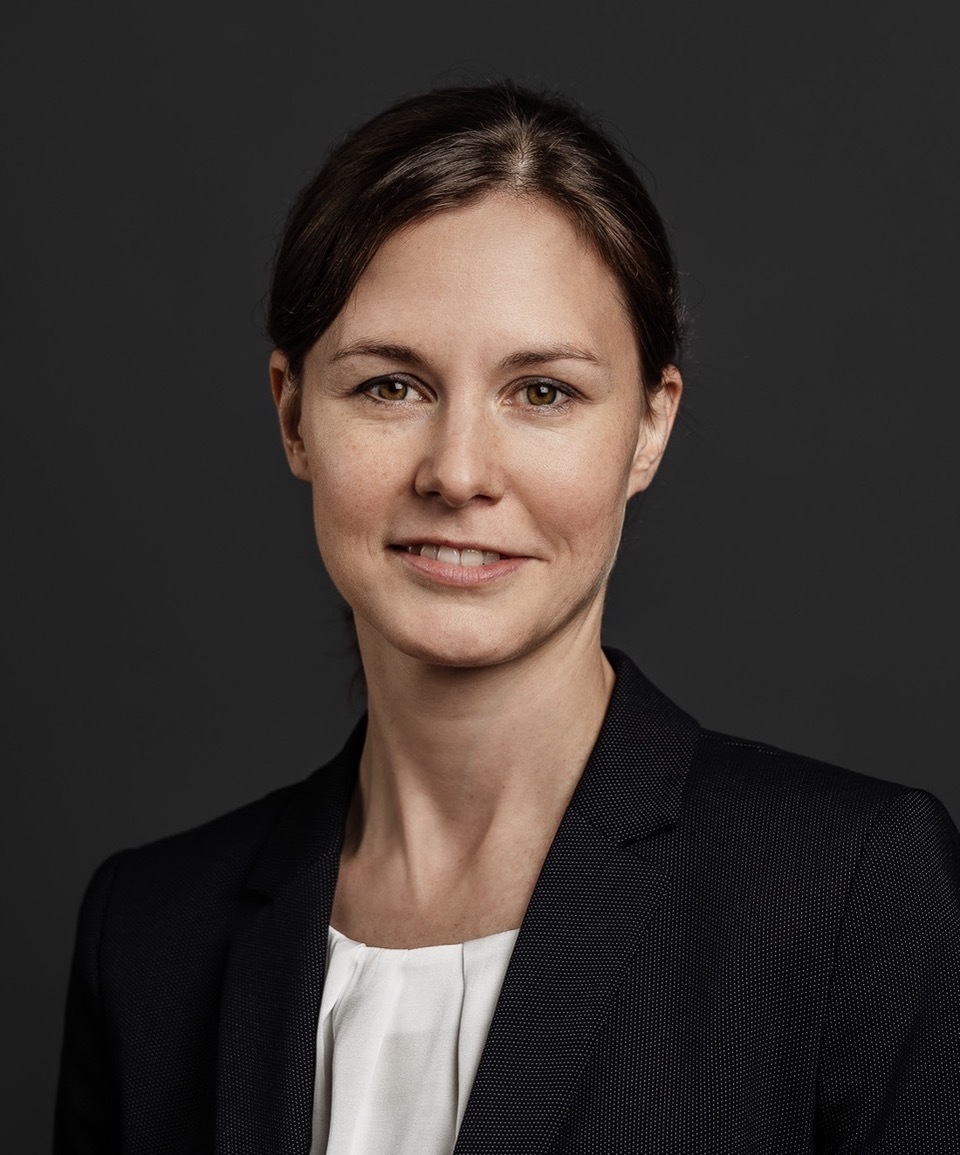}
\end{wrapfigure}
\noindent \textbf{Melanie N. Zeilinger} is an Associate Professor at ETH Zürich, Switzerland. She received the Diploma degree in engineering cybernetics from the University of Stuttgart, Germany, in 2006, and the Ph.D. degree with honors in electrical engineering from ETH Zürich, Switzerland, in 2011. From 2011 to 2012 she was a Postdoctoral Fellow with the Ecole Polytechnique Federale de Lausanne (EPFL), Switzerland. She was a Marie Curie Fellow and Postdoctoral Researcher with the Max Planck Institute for Intelligent Systems, Tübingen, Germany until 2015 and with the Department of Electrical Engineering and Computer Sciences at the University of California at Berkeley, CA, USA, from 2012 to 2014. From 2018 to 2019 she was a professor at the University of Freiburg, Germany. Her current research interests include safe learning-based control, with applications to robotics and human-in-the loop control.

\end{document}